\documentclass[aps,prl,twocolumn,superscriptaddress]{revtex4}

\usepackage{graphicx}
\usepackage{bbold}

\newcommand{\az}{\varphi}
\newcommand{\ro}{\rho}

\newcommand{\ga}{\gamma}

\newcommand{\oeq}{\begin{equation}}
\newcommand{\ceq}{\end{equation}}
\newcommand{\oeqn}{\begin{eqnarray}}
\newcommand{\ceqn}{\end{eqnarray}}

\renewcommand{\>}{\rangle}
\newcommand{\<}{\langle}

\renewcommand{\[}{\left[}
\renewcommand{\]}{\right]}
\newcommand{\lll}{\left|}
\newcommand{\rll}{\right|}

\newcommand{\oQ}{\hat{Q}}

\newcommand{\oX}{\hat{X}}
\newcommand{\oY}{\hat{Y}}

\newcommand{\oD}{\hat{D}}

\newcommand{\ovr}{\hat{\bf r}}

\newcommand{\oad}{\hat{a}^\dagger}
\newcommand{\oa}{\hat{a}}

\newcommand{\hb}{\hbar}

\renewcommand{\vr}{{\bf r}}

\newcommand{\Tr}{\mbox{Tr}}

\begin{document}

\title{Particle number fluctuations and correlations in transfer reactions obtained using the Balian-V\'en\'eroni variational principle}

\author{C\'{e}dric Simenel}\email[]{cedric.simenel@cea.fr}
\affiliation{Department of Nuclear Physics, Research School of Physics and Engineering, Australian National University, Canberra, Australian Capital Territory 0200, Australia}\affiliation{CEA, Centre de Saclay, IRFU/Service de Physique Nucl\'eaire, F-91191 Gif-sur-Yvette, France.}

\date{\today}

\begin{abstract}
The Balian-V\'en\'eroni (BV) variational principle, which optimizes the evolution of the state according to the relevant observable
in a given variational space, is used at the mean-field level 
to determine the  particle number fluctuations in fragments of many-body systems.
For fermions, the numerical evaluation of such fluctuations requires the use of a time-dependent Hartree-Fock (TDHF) code.
Proton, neutron and total nucleon number fluctuations 
in fragments produced in collisions of two $^{40}$Ca are computed for a large range of angular momenta 
at a center of mass energy $E_{c.m.}=128$~MeV, well above the fusion barrier. 
For deep-inelastic collisions, the fluctuations calculated from the BV variational principle
are much larger than standard TDHF results, 
and closer to mass and charge experimental fluctuations. 
For the first time, correlations between proton and neutron numbers are determined within a quantum microscopic approach.
These correlations are shown to be larger with exotic systems where charge equilibration occurs. 
\end{abstract}
\pacs{}
\maketitle

The quantum many-body problem is the root of many theoretical fields 
aiming at describing 
interacting particles 
such as electrons in metals, molecules, atomic clusters, Bose-Einstein condensates, or atomic nuclei~\cite{neg98}. 
However, it can be solved exactly for simple cases only. 
The Balian-V\'en\'eroni (BV) variational principle~\cite{bal81} 
offers an elegant starting point to build approximations of the many-body dynamics 
and has been applied to different problems 
in nuclear physics~\cite{tro85,mar85,bon85,zie88,bro08,bro09}, 
hot Fermi gas~\cite{mar91}, $\phi^4$ theory~\cite{mar95}, and Boson systems~\cite{ben99b,bou10}.
In particular, applications to deep-inelastic collisions (DIC) of atomic nuclei 
should be of interests to the upcoming exotic beam facilities. 
DIC will be used to investigate the role of the isospin degree of freedom in reactions
and to extract the density dependence of the symmetry energy.  
Such reactions will produce nuclei in extreme, sometimes unknown, states 
(e.g., rotating nuclei with a neutron skin, or nuclei at or beyond the drip-lines). 

Assuming an initial 
density matrix $\oD_0$ at $t_0$, 
The BV variational principle optimizes the expectation value of an observable $\<\oQ\>=\Tr(\oD\oQ)$ at a later time $t_1$. 
In this approach, both the state $\oD(t)$ and the observable $\oQ(t)$ vary between $t_0$ and $t_1$ 
within their respective variational spaces. 
In most practical applications, mean-field models are considered in a first approximation, 
and, eventually, serve as a basis for beyond-mean-field approaches~\cite{sim10a,lac04}. 
For instance, restricting the variational space of $\oD(t)$ to pure independent particle states, 
and the one of $\oQ(t)$ 
to one-body operators,  
leads to the TDHF equation~\cite{bal85}
\oeq 
i\hbar\frac{\partial \ro}{\partial t}=\[h[\ro],\ro\],
\label{eq:tdhf}
\ceq
where $\ro$ is the one-body density-matrix and $h[\ro]$ is the Hartree-Fock (HF) single-particle Hamiltonian.
According to this variational approach, TDHF is, then, the best mean-field theory 
to describe expectation values of one-body observables. 
However, it should not be used, in principle, 
to determine their fluctuations and correlations 
$
\sigma_{XY}=\sqrt{\<\oX\oY\>-\<\oX\>\<\oY\>},
\label{eq:corr_def}
$
($\oX$ and $\oY$ are one-body operators, and fluctuations correspond to the case $\oX=\oY$)
because the $\oX\oY$ term is outside the variational space of the observable.
Indeed, the TDHF expression for $\sigma_{XY}$,
\oeq
\sigma_{XY}^2(t_1)=\Tr \{Y\ro(t_1)X\[I-\ro(t_1)\]\},
\label{eq:corrTDHF}
\ceq
 where $I$ is the identity matrix, 
has been tested on fragment mass and charge fluctuations in DIC~\cite{koo77}.
In this case, $\oX=\oY$ counts the nucleons or protons of one fragment in the exit channel. 
$X$ and $Y$ are the matrices associated to $\oX$ and $\oY$, respectively, in single particle space.
 It was shown that TDHF strongly underestimates experimental fluctuations~\cite{koo77}. 
 This is an intrinsic limitation to TDHF~\cite{das79} which can be understood by the fact that, in such violent collisions, transfer of many particles may occur leading to very different mean-fields than the quasi-elastic one, while TDHF assumes that all mean-fields are the same. As a result, transfer of many particles are artificially hindered.
The knowledge of such fluctuations is, however, crucial to all quantum systems.
Thus, their theoretical prediction is an important challenge for quantum many-body models.

To optimize fluctuations of one-body operators, the variational space for the observable 
has to be increased to
$\oQ\in\{e^{\ga\oad\oa}\}$, where $\gamma\in\mathbb{R}$ and $\oa$ and $\oad$ are particle annihilators and creators, respectively~\cite{bal84,bal92}. This leads to a prescription
\oeq
\sigma_{XY}^2(t_1)=\lim_{\epsilon\rightarrow0}\frac{\Tr \{\[\ro(t_0)-\ro_X(t_0,\epsilon)\]\[\ro(t_0)-\ro_Y(t_0,\epsilon)\]\}}{2\epsilon^2}
\label{eq:corr}
\ceq 
differing from Eq.~(\ref{eq:corrTDHF}).
The one-body density matrices $\ro_X(t,\epsilon)$ obey 
the TDHF equation~(\ref{eq:tdhf}) with the boundary condition 
\oeq
\ro_X(t_1,\epsilon)=\exp({i\epsilon X})\ro(t_1)\exp({-i\epsilon X}),
\label{eq:boostrho}
\ceq
while $\ro(t)$ is the TDHF solution with initial condition $\ro(t_0)$ being the one-body density matrix of $\oD_0$. 
The result in Eq.~(\ref{eq:corr}) takes into account possible fluctuations around the TDHF mean-field evolution in the small amplitude limit, i.e., at the RPA level~\cite{bal84,bal92} [see also Ref.~\cite{ayi08} for an alternative derivation of Eq.~(\ref{eq:corr})].
These fluctuations are generated by the boost in Eq.~(\ref{eq:boostrho}) and propagated in the backward Heisenberg picture from $t_1$ to $t_0$ according to the dual of the time-dependent RPA equation.
This is why $\sigma_{XY}(t_1)$ is expressed as a function of density matrices at the initial time $t_0$. 
It is easy to show that, if the backward trajectories $\rho_X$ have the same mean-field as the forward evolution, then Eq.~(\ref{eq:corr}) leads to the TDHF expression in Eq.~(\ref{eq:corrTDHF}).
If, however, (small) deviations occur around the original mean-field, then additional terms appear and lead to an increase of $\sigma_{XY}(t_1)$.
In the following, TDHF and BV fluctuations or correlations refer to Eq.~(\ref{eq:corrTDHF}) and (\ref{eq:corr}), respectively.

In this work, the 
BV variational principle is used 
in realistic calculations
of heavy-ion collisions and first detailed comparisons with experiments are performed. 
The technique 
is similar to the one employed in~\cite{mar85,bon85}. 
The neutron, proton and mass fluctuations $\sigma_{NN}$, $\sigma_{ZZ}$, and $\sigma_{AA}$, respectively, 
and the correlation $\sigma_{NZ}$ between neutron and proton numbers distributions, 
are computed in fragments resulting both from deep-inelastic and quasi-elastic collisions. 
The correlations $\sigma_{NZ}$ 
are determined for the first time within a quantum microscopic approach. 


The {\textsc{tdhf3d}} code is used 
with the SLy4$d$ parameterization~\cite{kim97} of the Skyrme energy-density-functional~\cite{sky56}. 
The TDHF equation~(\ref{eq:tdhf}) is solved iteratively, 
with a time step~$\Delta{t}=1.5\times10^{-24}$~s, in the center of mass (c.m.) frame.
$\Delta{t}=10^{-24}$~s is also used to confirm the convergence of the BV fluctuations. 
The single-particle wave-functions are evolved on a Cartesian grid of $56\times56\times28/2$ points 
with a plane of symmetry (the collision plane) 
and a mesh-size $\Delta{x}=0.8$~fm. The initial distance between collision partners is 22.4~fm.
Refs.~\cite{sim10a,sim10b} give more details of the TDHF calculations.

To evaluate 
Eq.~(\ref{eq:corr}), the first step is to perform a TDHF evolution 
from $t_0$ to $t_1$.
To account for the transformation of Eq.~(\ref{eq:boostrho}), at time $t_1$, the occupied single particle wave functions are boosted according to 
$
|\az_{X_{j}}(t_1,\epsilon)\>=\exp(i\epsilon q_{X_j}\Theta(\ovr))|\az_{j}(t_1)\>,
$
where $X$ stands for $N$, $Z$, or $A$.
If the occupied single particle wave-function $\az_{j}$ refers to a proton (resp. a neutron), 
$q_{N_j}=0$ and $q_{Z_j}=1$ (resp. $q_{N_j}=1$ and $q_{Z_j}=0$), while $q_{A_j}=1$ for protons and neutrons. 
The function $\Theta(\vr)$ is equal to 1 for the fragment on which the fluctuations are calculated,
and 0 elsewhere. 
The time $t_1$ is determined, for each collision, by the time at which 
at least one fragment c.m. reaches 11.2~fm from one edge of the box.
It ensures a minimum separation distance of 22.4~fm for symmetric collisions. 
This value is large enough to ensure a convergence of $\sigma_{XY}$ with $t_1$ 
as the fragments interact only via Coulomb repulsion~\cite{bon85,mar85}. 

The second step is to compute a backward evolution from $t_1$ to $t_0$ 
of each set of single particle wave functions $\az_{X_{i}}(t,\epsilon)$.
Several values of $10^{-4}\le\epsilon\le10^{-2}$ are considered to determine the limit in Eq~(\ref{eq:corr}).
Following Refs.~\cite{bon85,bro09}, the initial density matrix $\ro(t_0)$ in Eq.~(\ref{eq:corr})
is replaced by a backward evolved density matrix $\ro_I(t_0,\epsilon=0)$, 
i.e., without the transformation in Eq.~(\ref{eq:boostrho}), to minimize numerical inaccuracies.
Note that the latter are easily controllable and this procedure is not necessary with a smaller time step $\Delta{t}$.
The trace in Eq.~(\ref{eq:corr}) is then evaluated with
$\Tr \{\[\ro_I(t_0,0)-\ro_X(t_0,\epsilon)\]\[\ro_I(t_0,0)-\ro_Y(t_0,\epsilon)\]\} =\eta_{II}+\eta_{XY}-\eta_{IX}-\eta_{IY},
$ where $\eta_{XX'}=\sum_{ij}\lll\<\az_{X_i}(t_0)|\az_{X'_j}(t_0)\>\rll^2$, and the sums run over occupied states.
The quadratic evolution of the trace with $\epsilon$ is used as a convergence check~\cite{mar85}, 
as well as the property $\eta_{II}=A_t$, where $A_t$ is the total number of nucleons.

\begin{figure}
\includegraphics[width=7.5cm]{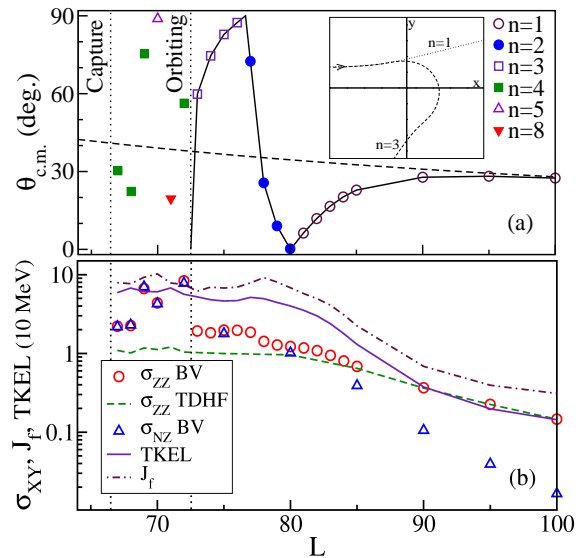} 
\caption{(a) Scattering angle as a function of angular momentum $L$ 
in units of~$\hb$ for $^{40}$Ca+$^{40}$Ca at $E_{c.m.}=128$~MeV. 
$n$~is the number of times the $x$ (collision axis) or $y$ axis has been crossed by the fragment. 
The solid line is to guide the eye. The dashed line corresponds to Rutherford trajectories. 
The inset gives examples of trajectories for $n=1$ and 3. 
(b) TDHF (dashed line) and BV (circles) fluctuations of $Z$, 
BV correlations between $N$ and $Z$ (triangles), 
and intrinsic angular momentum $J_f$ of the outgoing fragments (dot-dashed line).
The TKEL (solid line) is given in units of 10 MeV. 
The vertical dotted lines show the range of $L$ leading to orbiting 
or capture trajectories.
}
\label{fig:sigma}
\end{figure}

Let us first investigate the collision of two $^{40}$Ca nuclei at $E_{c.m.}=128$~MeV. 
This system is considered as a benchmark for experimental studies of DIC~\cite{roy77,eva91}. 
Theoretically, the HF ground-state of $^{40}$Ca is expected to be a good approximation because of its doubly-magic nature.
Fig.~\ref{fig:sigma}(a) shows the scattering angle $\theta_{c.m.}$ 
as a function of the angular momentum $L$.
For $L\ge90$, the trajectories are close to Rutherford scattering (dashed line) and correspond to quasi-elastic reactions 
with small total kinetic energy loss (TKEL) as shown by the solid line in Fig.~\ref{fig:sigma}(b).
At lower $L$, deviations from the Rutherford formula occur because of nuclear attraction, 
leading to rotation of the di-nuclear system formed by the two fragments in contact. 
For instance, $L\le72$ defines (arbitrarily) the orbiting region where the fragments have crossed their incoming trajectory, 
i.e., with $n\ge4$ crossings of the $x$ (collision axis) or $y-$axis [see inset in Fig.~\ref{fig:sigma}(a)]. 
Following~\cite{roy77}, damped events are defined by a TKEL$\ge30$~MeV, 
corresponding to $L<82$ in Fig.~\ref{fig:sigma}. 
We see that a wide range of scattering angles may occur for these damped events, 
which is a known feature of DIC. 
For $L\le66$, capture occurs. It corresponds to a fusion cross section of $\sim1140$~mb with the sharp cutoff formula~\cite{sim10a}, in good agreement with a fit on fusion-evaporation measurements at lower energies~\cite{dou78}.

Fig.~\ref{fig:sigma}(b) shows BV and TDHF predictions of $\sigma_{ZZ}$ (only charge fluctuations are shown for clarity). 
The TDHF fluctuations [Eq.~(\ref{eq:corrTDHF})] have been determined from the probability distributions 
of $A$, $Z$ and $N$ in the fragments at time $t_1$~\cite{sim10b}.  
BV predictions from Eq.~(\ref{eq:corr}) at $L=71$ are not shown as no numerical convergence with $\epsilon$ could be obtained. 
Strong variations of the BV fluctuations are observed in the orbiting region, 
but there is no clear relationship with the amount of orbiting quantified by $n$ in Fig.~\ref{fig:sigma}(a).
The BV fluctuations are much more important than the TDHF predictions for damped events.
However, at large $L$ (quasi-elastic reactions), the BV and TDHF fluctuations are similar. 

\begin{figure}
\includegraphics[width=6.5cm]{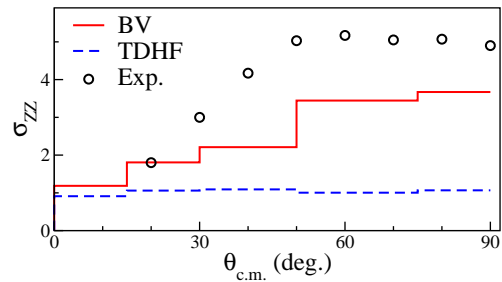} 
\caption{Comparison between BV (solid line) and TDHF (dashed line) predictions of $\sigma_{ZZ}$ for damped events (see text) as a function of $\theta_{c.m.}$ with data (circles) from~\cite{roy77}.}
\label{fig:angle}
\end{figure}

The evolution of $\sigma_{ZZ}$ with $\theta_{c.m.}$ is plotted in Fig.~\ref{fig:angle} for damped events, 
and compared with experimental data~\cite{roy77}. 
TDHF fluctuations show no angular correlation and strongly underestimate data.
BV fluctuations increase with $\theta_{c.m.}$ at small angles and form a plateau at large angles, in qualitative agreement with data. 
Quantitatively, the experimental plateau is underestimated. 
This might be attributed to fusion-fission events leading to large fluctuations and not included in the calculations. 
Indeed, the compound-nucleus fission cross-section, estimated to be $\sim280$~mb~\cite{roy77}, 
is not negligible compared to the cross-section for damped binary events of~$\sim570$~mb from the present calculations.  
Note that this fission cross-section seems reasonable as it corresponds to a fusion-evaporation cross-section of $\sim860$~mb
(obtained from the difference between the fusion and fusion-fission cross sections), which is compatible with data~\cite{dou78}.
Due to the isotropic distribution of fission fragments,  
fusion-fission would mostly affect large angles in Fig.~\ref{fig:angle} 
and may account for the difference between BV predictions and data. 
As the fragments cool down by nucleon emission, their fluctuations might also increase~\cite{bro08}.
Although the number of TDHF iterations is too small to allow a full decay of the fragments by nucleon emission,
we can 
estimate their excitation energy 
$E^*\simeq{TKEL}/2$ and angular momenta $J_f\simeq(L-L_{out})/2$ [see Fig.~\ref{fig:sigma}(b)], 
assuming equal sharing, where $L_{out}$ is the angular momentum between the fragments in the outgoing channel. 
Calculations using the code {\textsc{PACE4}}~\cite{gav80} with a level density parameter $A/7.5$~MeV$^{-1}$ 
show that the decay of fragments produced in DIC has only a small effect on $\sigma_{ZZ}$ (see also~\cite{koo77}).
However, the average of the fragment charge distribution after decay goes from $\bar{Z}\simeq19$ for $L\simeq80$ 
down to 18 in the orbiting region while experimental data give $\bar{Z}\simeq17$ at large angle~\cite{roy77}.
This might also be a signature of fusion-fission events. 
Indeed, symmetric fission leads to $E^*\simeq38.2$~MeV according to the Viola systematics~\cite{vio85}, and, then,
to more emission of light particles than in DIC where $E^*\simeq31$~MeV 
in average in the orbiting region [see Fig.~\ref{fig:sigma}(b)].
Beyond mean-field correlations may also affect these fluctuations. 



Fragment mass distributions in $^{40}$Ca$+^{40}$Ca have been measured at lower energies, 
$E_{c.m.}=98.5$ and 115.5~MeV, where fusion-fission can be neglected~\cite{eva91}.
In this experiment, the fragments are associated to almost fully damped collisions 
with a $\sin^{-1}\theta_{c.m.}$ dependence of their cross-section. 
In the present work, such collisions occur in the orbiting region for which $\sigma_{AA}\simeq9.7$ in average.  
This is in fair agreement with the data which give $\sigma_{AA}\simeq11$.


In addition to fluctuations, the BV correlations $\sigma_{NZ}$ 
have been computed 
[triangles in Fig.~\ref{fig:sigma}(b)]. 
These finite values of $\sigma_{NZ}$ are at variance with the TDHF correlations which are strictly zero
because the single particle states are assumed to have pure isospin. 
In fact, the probability $P(N,Z)$ to have a fragment with $Z$ and $N$, in TDHF calculations, 
is the product of the probabilities $P(Z)P(N)$ to have $Z$ and $N$, independently~\cite{sim10b}.
For instance, in the symmetric collisions studied here, the TDHF probability to have the $N=Z$ $^{32}$S
nucleus is the same as for the neutron rich $^{40}$S.
The latter should, however, be hindered by the symmetry energy 
which induces a fast charge equilibration in the fragments~\cite{iwa10}.
Such effect is included in the BV approach which give $\sigma_{NZ}\simeq\sigma_{ZZ}$ 
in damped collisions, 
while, in quasi-elastic reactions, correlations are negligible compared to fluctuations [see Fig.~\ref{fig:sigma}(b)]. 
Note that the fact that BV and TDHF predictions are similar for large $L$ allows the use of TDHF to investigate quasi-elastic reactions~\cite{sim10b}.  
The present calculations also give $\sigma_{NN}\simeq\sigma_{ZZ}$ for all $L$, $\sigma_{AA}\simeq1.5\sigma_{ZZ}$ for $L\ge90$ and  $\sigma_{AA}\simeq2\sigma_{ZZ}$ in the orbiting region. This is in good agreement with 
the relation $\sigma_{AA}^2=\sigma_{ZZ}^2+\sigma_{NN}^2+2\sigma_{NZ}^2$ and the behavior of $\sigma_{NZ}$.

Finally, the $^{80,92}$Kr+$^{90}$Zr systems at beam energy $E/A=8.5$~MeV 
are investigated to study the role of isospin asymmetry in the entrance channel. 
The calculations are performed at $L=192.2$ with $^{80}$Kr and 222.6 with $^{92}$Kr 
corresponding to touching spheres of radii $1.2A^{1/3}$ at closest approach.
The $N/Z$ ratio of $^{80}$Kr and $^{90}$Zr are similar (1.22 and 1.25, respectively) but differ from the one of $^{92}$Kr (1.56). 
Then, charge equilibration occurs only with $^{92}$Kr 
where the calculations give an average $N/Z\simeq1.4$ in both outgoing fragments.
The BV prescription gives $\sigma_{ZZ}\simeq5.3$, $\sigma_{NN}\simeq7.1$, and $\sigma_{NZ}\simeq5.7$ with $^{80}$Kr, 
and $\sigma_{ZZ}\simeq4.7$, $\sigma_{NN}\simeq8.4$, and $\sigma_{NZ}\simeq8.5$ with $^{92}$Kr.
Fluctuations are only slightly affected, while charge equilibration strongly increases correlations between $N$ and $Z$ distributions (by $\sim50\%$).


The BV prescription leads to fragment mass and charge fluctuations in better agreement with experiment 
than TDHF for violent collisions, although TDHF should be sufficient for quasi-elastic reactions. 
The predictions of correlations between $N$ and $Z$ distributions 
is an attractive feature which should
be compared with experimental data where both mass and charge of each fragment are measured. 
In particular, reactions with exotic beams will allow strong $N/Z$ asymmetries, 
and such correlations are expected to be increased by the charge equilibration process. 
Applications to multi-nucleon transfer in actinide collisions could be used to predict
probabilities for super-heavy element production~\cite{gol09,ked10}.
The role of pairing correlations on fluctuations should be investigated using
recent 
time-dependent Hartree-Fock-Bogoliubov codes~\cite{ave08,eba10,was10}.
Stochastic-mean-field methods might also be applied to investigate the role of initial beyond-mean-field correlations on 
fluctuations~\cite{ayi08,was09b}. 

\begin{acknowledgments}
M.~Dasgupta, 
D.~J.~Hinde, and D. Lacroix are thanked for discussions. 
The calculations have been performed
on the NCI National Facility 
in Canberra, Australia, which is supported by the Australian 
Commonwealth Government.
The author acknowledges the support of ARC Discovery grant DP 0879679.
\end{acknowledgments}

\end{document}